\documentclass[%
 reprint,
superscriptaddress,
 amsmath,amssymb,
 aps,
longbibliography
]{revtex4-1}
\usepackage{xcolor}
\usepackage{graphicx}
\usepackage{dcolumn}
\usepackage{bm}

\begin{document}

\preprint{APS/123-QED}

\title{Effect of Interfacial Thermal Resistance in Thermal Cloak}

\author{Xu Zheng}
\affiliation{Department of Physics, University of Colorado, Boulder, CO, 80309, USA}
\author{Baowen Li}
\email{Baowen.Li@Colorado.Edu}
\affiliation{Department of Mechanical Engineering, University of Colorado, Boulder, CO, 80309, USA}
\affiliation{Department of Physics, University of Colorado, Boulder, CO, 80309, USA}
\homepage{https://www.colorado.edu/faculty/li-baowen/}
\date{\today}

\begin{abstract}
 When heat transfers through interface between two different materials, it will encounter an interfacial thermal resistance (ITR) that makes the temperature discontinuous. This effect has been totally neglected   so far in the research of thermal cloak, in particular when the thermal cloak is built with multilayer structures. In this paper, we investigate the effect of ITR on the performance of the thermal cloak  by using both analytical and numerical method. Our results show that the existence of ITR will distort the external field, thus destroy the cloak. Moreover, we found that the effect of ITR can be quantified by a parameter called  characteristic length. 
\end{abstract}

\maketitle


\section{Introduction}

Management and control of heat transfer such as refrigerating/air-conditioning and converting into electricity, has always been one of the most important issues not only in our daily life but also in further technology development \cite{segre2003matter}.  Indeed, heat death - excess heat generation breaks down electronic devices - has been an outstanding problem for further development of microelectronics \cite{waldrop2016chips}.
Recent years have witnessed an important progress in developing novel heat control methods. At the microscopic level, in analog with electronics, a new fast-growing field called phononics \cite{li2012colloquium} has been emerged to manipulate heat conduction due to phonons in semiconductors and dielectric materials. At the macroscopic level, thermal metamaterials have been used to control both heat radiation \cite{fan2018metamaterials,zhai2017scalable} and heat conduction \cite{SophiaLi}. The concept of thermal metamaterials for manipulating heat conduction or sometimes also called transforming heat conduction \cite{guenneau2012transformation} was inspired by the idea of transformation optics \cite{pendry2006controlling,schurig2006metamaterial},
which provides a general method to realize complex control of heat flux, among which thermal cloak is of special interest. 

Thermal cloak is a structured material that any object concealed inside this material will not alter the temperature and heat flux distribution outside the cloak. Therefore, two criteria are important in evaluating the performance of a thermal cloak: (i) the heat flux in the cloaking region should be as small as possible; (ii) the temperature distribution outside the cloak should be the same as the pure background to realize invisibility. 

The general heat conduction equation is of the form

\begin{equation}
    c(\bm{x})\rho(\bm{x})\frac{\partial T}{\partial t}=\nabla\cdot(\kappa(\bm{x})\nabla T)+Q(\bm{x})
\end{equation}
where $T$ represents the temperature distribution evolving with time, $c(\bm{x})$ is the specific heat capacity, $\rho(\bm{x})$ is the volumetric mass density, $\kappa(\bm{x})$ is the thermal conductivity, and $Q(\bm{x})$ is the internal heat source term. 

Based on this equation, Guenneau et al derive a rigorous transformation formula to design a perfect thermal cloak by using inhomogeneous and continuous parameters, i.e., $c(\bm{x})\rho(\bm{x})$ and $\kappa(\bm{x})$ \cite{guenneau2012transformation}.
To make things simple, people often consider the heat conduction in steady state without internal heat sources. In this case, the heat conduction equation can be simplified to
\begin{equation}
    \nabla\cdot(\kappa(\bm{x})\nabla T)=0.
\end{equation}

Mathematically, this equation is the same as the steady-state wave equation. This indeed has inspired Fan et al \cite{fan2008shaped} to study steady thermal cloak before the general consideration given by ref. \cite{guenneau2012transformation}. As a special case of the general heat conduction equation, only one continuous tensor $\kappa(\bm{x})$ is required. 

However, practically it is still nearly impossible to realize continuous thermal conductivity $\kappa(\bm{x})$. As a good approximation, multilayer structures have been used to realize thermal cloaks \cite{narayana2012heat,schittny2013experiments,xu2014ultrathin,ma2014experimental,han2014experimental}. Recently, Choe et al even realized a multilayer thermal cloak at the mircoscale by using helium ion beam writing technique \cite{choe2019ion}. However, any attempt to use multilayer structures to approximate the continuous thermal cloak will introduce the Interfacial Thermal Resistance (ITR) or called Kapitza resistance \cite{kapitza1941study,pollack1969kapitza,swartz1989thermal,giri2019review}, which has been ignored in all of those experiments and theoretical/numerical analysis. 

In general, heat transfer through boundary between two different materials will encounter ITR and the effect is not negligible \cite{kapitza1941study,pollack1969kapitza,swartz1989thermal,giri2019review}. Moreover, at the nanoscale, the ITR becomes asymmetric and leads to thermal rectification \cite{li2005interface}. ITR is the combination of two resistances: (i) thermal contact resistance (TCR) due to poor mechanical contact and (ii) thermal boundary resistance (TBR) due to differences in physical properties in different materials. 

Currently, the boundary condition used in the design of multilayer thermal cloak is continuous condition, i.e.,
\begin{align}
    &T_{i+1}|_{\partial\Omega}=T_{i}|_{\partial\Omega}\nonumber\\
    & \bm{n}\cdot\left(\kappa_i\nabla T_{i}\right)|_{\partial\Omega}=\bm{n}\cdot\left(\kappa_{i+1}\nabla T_{i+1}\right)|_{\partial\Omega}
\end{align}
where $T_i$, $T_{i+1}$ is the temperature of the $i$th, $(i+1)$th layer respectively. $\partial\Omega$ represents the interface between the $i$th and $(i+1)$th layers, $\bm{n}$ is the unit normal vector of the interface, and $\kappa_i$, $\kappa_{i+1}$ is the thermal conductivity of the $i$th, $(i+1)$th layer, respectively. When ITR is considered, the temperature continuity is broken while the continuity of heat flux is still true, there exists a temperature drop at the interface
\cite{kapitza1941study,pollack1969kapitza,swartz1989thermal}

 \begin{align}
     &T_{i+1}|_{\partial\Omega}=T_{i}|_{\partial\Omega}+R_{i,i+1}\bm{n}\cdot\left(\kappa_i\nabla T_{i}\right)|_{\partial\Omega}\nonumber\\
     & \bm{n}\cdot\left(\kappa_i\nabla T_{i}\right)|_{\partial\Omega}=\bm{n}\cdot\left(\kappa_{i+1}\nabla T_{i+1}\right)|_{\partial\Omega}
 \end{align}
 where $R_{i,i+1}$ is the ITR between the $i$th and $(i+1)$th layer . As the thermal cloak and some other thermal devices are coming to nanoscale, we should notice that ITR can become significant. Therefore, it is quite natural to understand the effect of ITR in thermal cloaking. 

In this paper, we investigate the effect of ITR, mainly the ITR of cloak/background interface and cloak/cloaked object interface, in the heat flux and temperature distribution, and further propose modifications to minimize and/or get rid of this effect. In particular, we focus on the two-dimensional (2D) cloak.
The paper is organized as the following: in Section II, we discuss the effect on a 2D anisotropic single layer cloak; in Section III, we discuss the case for 2D bilayer cloak; and finally in the Section IV we give discussions.

\section{Anisotropic Single Layer Cloak}
\begin{figure}[htbp]
    \centering
    \includegraphics[width=0.6\linewidth]{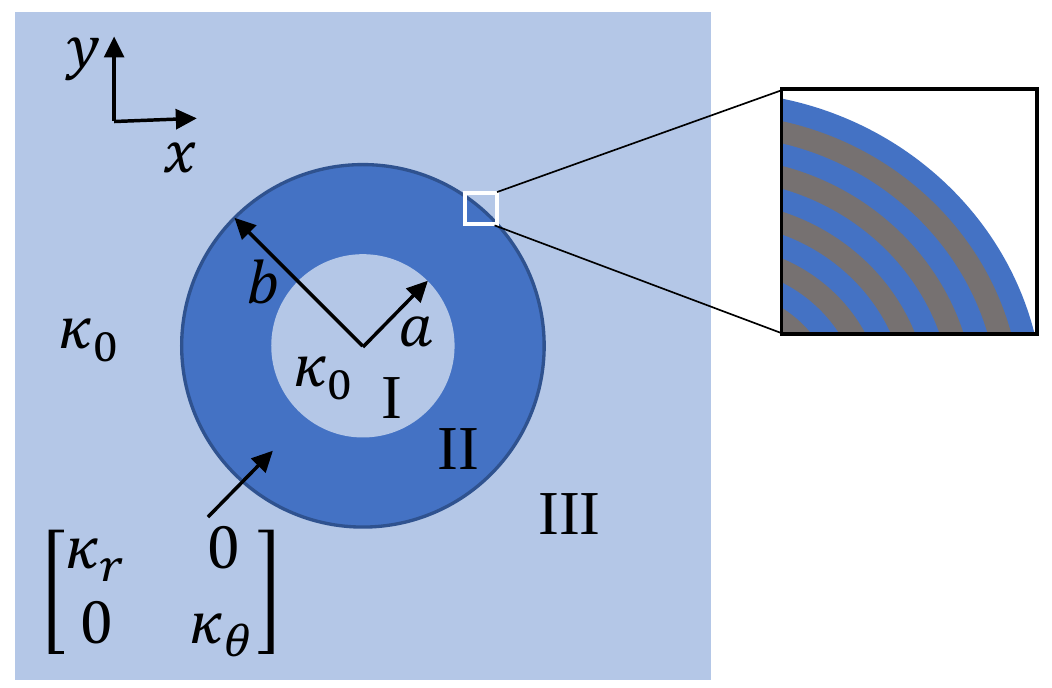}
    \caption{Schematic of anisotropic single layer cloak.}
    \label{fig:struct}
\end{figure}

First, we consider a 2D single layer structure shown in Fig. (1). The annular cloak (region II) has a diagonal thermal conductivity that satisfies $\kappa_r\kappa_{\theta}=\kappa_0^2$ in polar coordinate, while the cloaking object (region I) and background (region III) is isotropic and homogeneous with thermal conductivity $\kappa_0$. The whole structure is presented in a uniform thermal gradient field $\nabla T_b=G_0\hat{x}$. This cloak can be realized through annular multilayered composition consisting of two different isotropic materials (medium A and medium B) arranged alternately, as shown in the inset of Fig. (1). If the thermal conductivity of medium A (B) is $\kappa_A$ ($\kappa_B$), layer thickness is $d_A$ ($d_B$), and the ITR is $R_{AB}$. The effective thermal conductivity of the composition is \cite{simkin2000minimum}
\begin{align}
    \frac{d_A+d_B}{\kappa_{r,eff}}=\frac{d_A}{\kappa_A}+\frac{d_B}{\kappa_B}+2R_{AB}\\
    (d_A+d_B)\kappa_{\theta,eff}=d_A\kappa_A+d_B\kappa_B
\end{align}
The case of no ITR has been well studied by Han et al \cite{han2013homogeneous}. They show that nearly perfect cloaking performance can be achieved as long as $\kappa_r/\kappa_0=\kappa_0/\kappa_{\theta}$ is smaller than $\frac{\lg{(a/b)}}{\lg{(a/b)}-1}$. The physical picture is that when $\kappa_r\ll\kappa_{\theta}$, the heat flux tends to align along the tangential direction to bypass the cloaking region. In the case of considering ITR, the general solutions of temperature distribution in polar coordinate can still be expressed as \cite{han2013homogeneous}
\begin{align}
T_1&=\sum_{n=1}^{\infty}A_{n}r^{n}\cos{n\theta}\\
T_2&=\sum_{n=1}^{\infty}\left[B_{n}r^{nl}+C_{n}r^{-nl}\right]\cos{n\theta}\\
T_3&=\sum_{n=1}^{\infty}\left[D_{n}r^{n}+E_{n}r^{-n}\right]\cos{n\theta}
\end{align}
where $l=\sqrt{\kappa_{\theta}/\kappa_r}$, $T_{1,2,3}$ denotes the temperature distribution is region I, II, II respectively, and $A_n$, $B_n$, $C_n$, $D_n$, $E_n$ are constants determined by boundary conditions. At infinity, the background thermal gradient field is not affected by the cloak, i.e., 
\begin{align}
    &T_3|_{r\rightarrow\infty}=-G_0r\cos{\theta},
\end{align}
Combining Eq. (10) and the boundary conditions Eq. (4) at the interface I/II and II/III, the solutions are 
\begin{align}
    A_1&=-G_0\left(\frac{a}{b}\right)^{l-1}\frac{1}{F}\\
    D_1&=-G_0\\
    E_1&=-\frac{G_0b^2H}{F}
\end{align}
where 
\begin{align}
    &H=\frac{R_{23}\kappa_0}{2b}+\frac{R_{12}\kappa_0}{2b}\frac{a^{2l-1}}{b^{2l-1}}+\frac{R_{12}R_{23}\kappa_0^2}{4ab}\left(1-\frac{a^{2l}}{b^{2l}}\right)\\
    &F=1+\frac{R_{12}\kappa_0}{2a}+\frac{R_{23}\kappa_0}{2b}+\frac{R_{12}R_{23}\kappa_0^2}{4ab}\left(1-\frac{a^{2l}}{b^{2l}}\right)
\end{align}
and all other $A_{n}$'s, $D_{n}$'s and $E_{n}$'s are zero. $R_{12}$ ($R_{23}$) denotes the ITR of interface I/II (II/III). $A_1$ evaluates the heat flux in cloaking region and $E_1$ evaluates the field distortion outside the cloak. We are only interested in cloaking region (I) and external region (III), so we don't explicitly give the solutions in region II. Let's assume $R_{12}=R_{23}=R_I$ since the material of region I and III are of the same in current structure. If there is no ITR ($R_I=0$, $F=1$, $H=0$), these solutions are the same as those in Ref. \cite{han2013homogeneous}. 

In region I, the effect of ITR is introducing a scale factor $1/F$. Since $F\geq 1$, $1/F\leq 1$. The heat flux in region I becomes smaller when considering ITR, which means better thermal shielding. We can further investigate the change of scale factor $1/F$ with respect to $l=\sqrt{\kappa_{\theta}/\kappa_r}$ and $a$, $b$. 
The ITR is usually in the range $10^{-9}\sim10^{-7}$ K$\cdot$m$^2$/W \cite{giri2019review}. Here we assume $R_I=1\times10^{-7}$ K$\cdot$m$^2$/W, the scale factor $1/F$ as a function of $l$ and $a$, $b$ is shown by blue solid lines in Fig. 2. At fixed $a$ and $b$, the scale factor approaches a finite value as $l$ increases. At fixed $l$, the scale factor approaches zero as the cloak size decrease.

In region III, thermal cloak requires that there is no external-field distortion, which means $E_1$ should be zero. But from Eq. (13)-(15), we find $E_1<0$ as long as there is ITR (notice that the factor $H$ and $F$ is always greater than zero) . So the performance of thermal cloak is worse with respect to the external-field distortion when considering ITR. To quantify the distortion, we plot the ratio of temperature difference ($\Delta T=T_3-T_b$) induced by the ITR to the background thermal gradient field ($T_b$) at a point ($b,0$), as shown by red dash lines in Fig. (2). A direct calculation gives $\Delta T/T_b=H/F$ at point ($b,0$). At fixed cloak size $a$ and $b$, Fig.2 (a) shows the distortion decreases with increasing $l$, but there exists a lower bound. At fixed $l$, Fig. 2(b) shows that the distortion quickly increases as the cloak size decreases.

An interesting feature of the factor $F$ and $H$ is that they define a characteristic length
\begin{equation}
    L_c=R_I\kappa_0
\end{equation}
which is the product of ITR and background thermal conductivity. When the characteristic length $L_c$ is comparable to, or even greater than, the cloak size $a$ and $b$, both the reduction of heat flux in region I and external-field distortion in region III will become significant. As shown in Fig. 2(b), when cloak size $b$ equals to characteristic length $L_c$, the scale factor $1/F$ can be as small as 0.3, and $\Delta T/T_b$ can be larger than 0.3.

\begin{figure}[htbp]
    \centering
    \includegraphics[width=\linewidth]{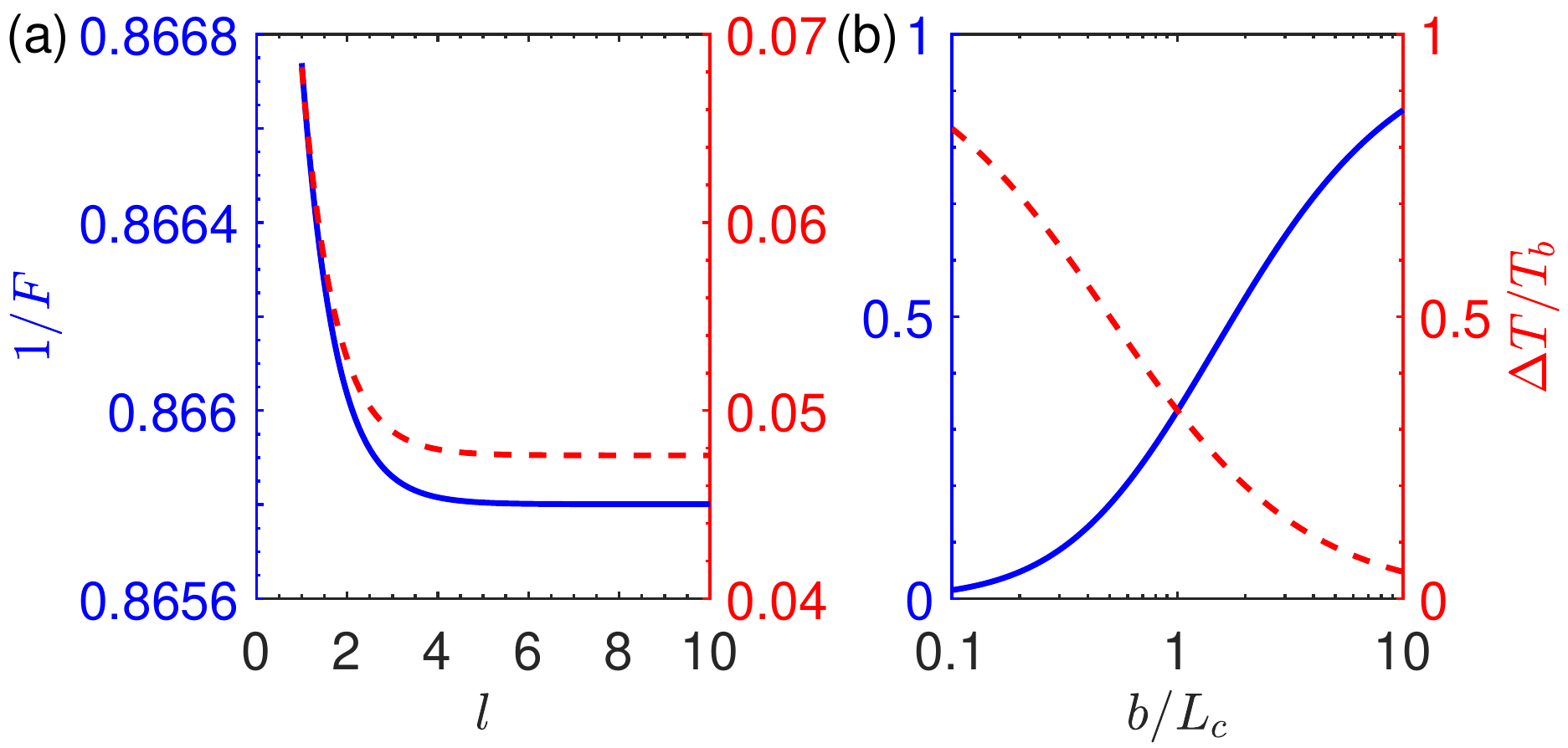}
    \caption{Effect of ITR as a function of $l$ and $b$. $1/F$ is marked by blue solid lines and $\Delta T/T_b$ is marked by red dash lines. (a) $1/F$, $\Delta T/T_b$ as a function of $l$. $a=0.5$ $\mu$m, $b=1$ $\mu$m. (b) $1/F$, $\Delta T/T_b$ as a function of cloak size $b$. $l=10$, the ratio $a/b$ is fixed to 0.5. In (a) and (b), $\kappa_0=1$ W/(m$\cdot$K), $R_I=1\times10^{-7}$ K$\cdot$m$^2$/W, $L_c=R_I\kappa_0=1\times10^{-7}$ m. }
    \label{fig:fgvsl}
\end{figure}

\begin{figure}
\centering
    \includegraphics[width=0.9\linewidth]{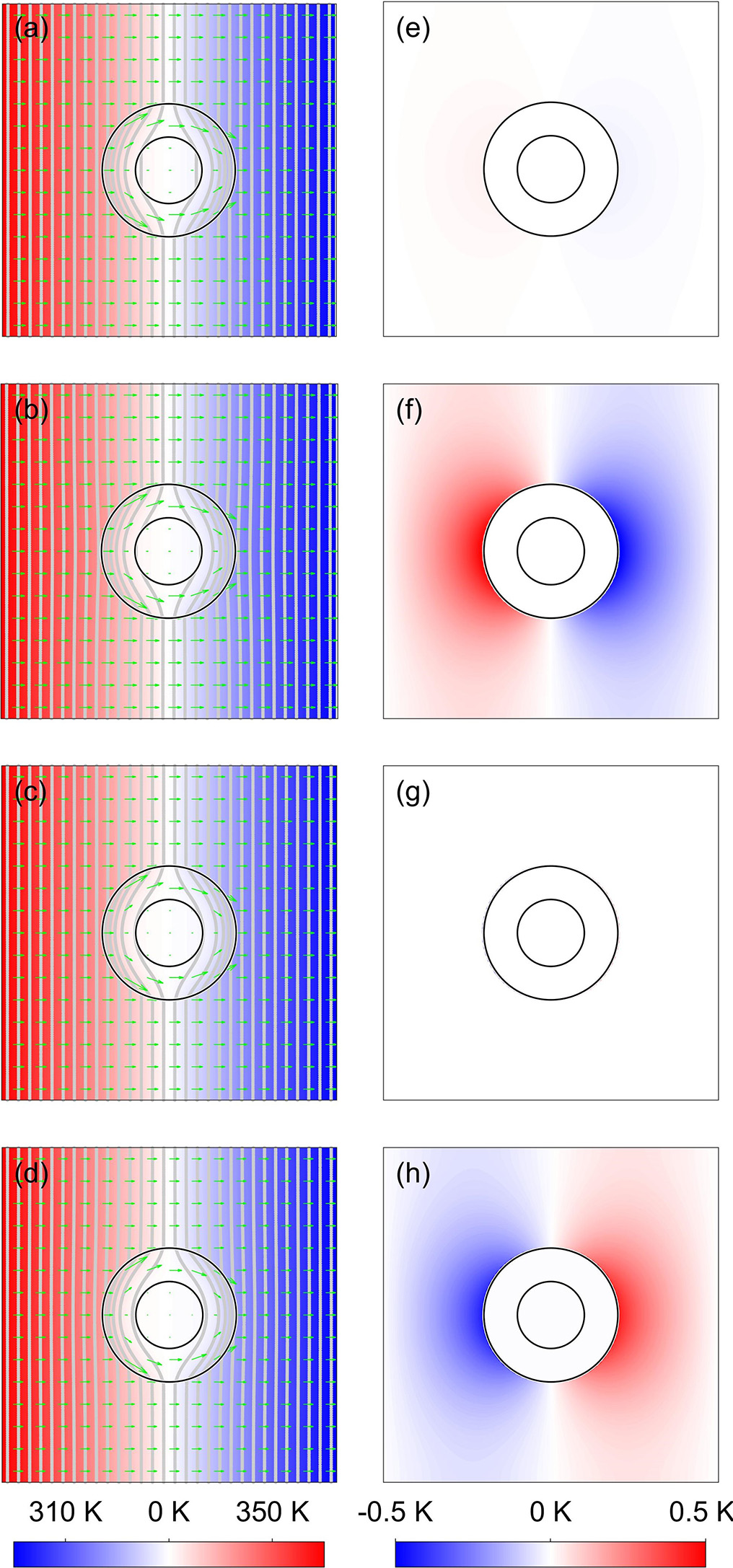}
\caption{Temperature distribution of the single layer structure. The green arrows show the heat flux distribution. Temperature at the left (right) boundary is fixed at 360 K ($300$ K). $a=0.5$ $\mu$m, $b=1$ $\mu$m, $\kappa_0=1$ W/(m$\cdot$K), $\kappa_r=0.3$ W/(m$\cdot$K). (a) $\kappa_{\theta}=3.3$ W/(m$\cdot$K), $R_I=0$ K$\cdot$m$^2$/W. (b) $\kappa_{\theta}=3.3$ W/(m$\cdot$K), $R_I=1\times10^{-7}$ K$\cdot$m$^2$/W. (c) $\kappa_{\theta}=4.13$ W/(m$\cdot$K), $R_I=1\times10^{-7}$ K$\cdot$m$^2$/W. (d) $\kappa_{\theta}=5$ W/(m$\cdot$K), $R_I=1\times10^{-7}$ K$\cdot$m$^2$/W. (e)-(h) Temperature difference between the temperature distribution in (a)-(d) and the uniform thermal gradient field. Only the difference outside the cloak region is calculated because we want to study the external-field distortion.}
\label{fig:singcol} 
\end{figure}

\begin{figure}
    \centering
    \includegraphics[width=\linewidth]{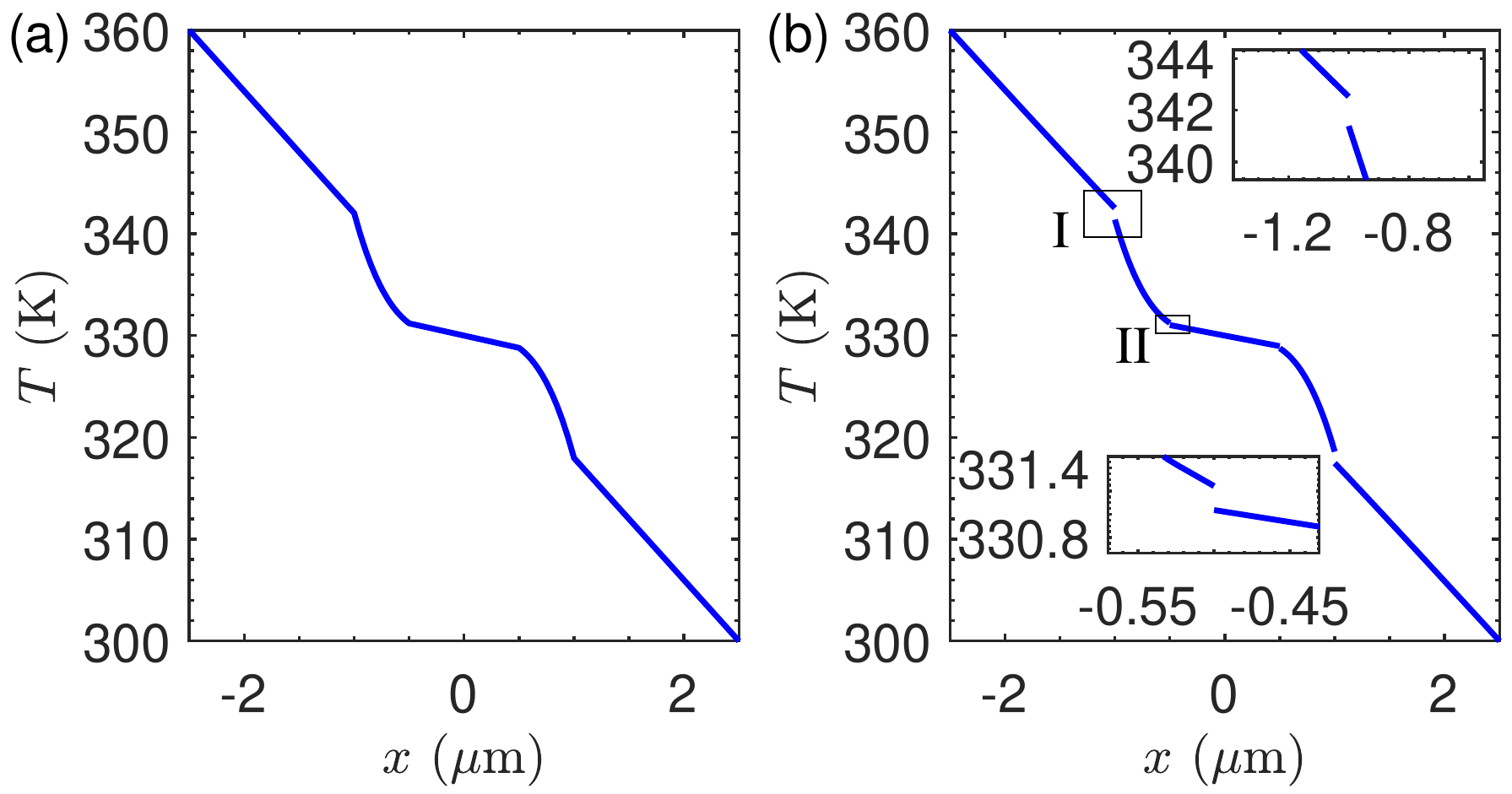}
    \caption{Temperature distribution along $x$ axis. Parameters of Fig. 3(a) and 3(b) are used in (a) and (b) respectively. The upper right (lower left) inset shows the zoomed rectangle I (II). Discontinuity at the interface $a=\pm 0.5$ $\mu$m and $b=\pm 1$ $\mu$m can be clearly observed when considering ITR.}
    \label{fig:tempalongx}
\end{figure}

Fig. 3 shows the numerical simulation of temperature distribution and heat flux based on finite element method (FEM). In Fig. 3(a) and 3(b), we can find that the heat flux in region I is smaller when considering ITR. By calculating the temperature difference with respect to the uniform thermal gradient field, it is clear that the external field is distorted in region III when considering ITR, as shown in Fig. 3(b) and 3(f). 

Fig. 4 shows the temperature distribution of Fig. 3(a) and 3(b) along the $x$ axis. The discontinuity is obvious at the interface $a=0.5$ $\mu$m and $b=1$ $\mu$m in the existence of ITR. 

Next, we would like to know if we can eliminate the external-field distortion by modifying thermal conductivity. A direct generalization of the original idea is using $\kappa_r\kappa_{\theta}=t^2\kappa_0^2$, where $t$ is a constant to be determined, instead of $\kappa_r\kappa_{\theta}=\kappa_0^2$. The new solutions of heat conduction equation hold the same forms of Eq. (11)-(13) except that now the factors $H$ and $F$ are functions of $t$,
\begin{align}
    &F(t)=\frac{1}{4t}\left[1-\frac{a^{2l}}{b^{2l}}+2\left(1+\frac{a^{2l}}{b^{2l}}\right)\left(1+\frac{R_I\kappa_0}{2a}+\frac{R_I\kappa_0}{2b}\right)t\right. \nonumber\\ 
    &~~~~~~+\left.\left(1-\frac{a^{2l}}{b^{2l}}\right)\left(1+\frac{R_I\kappa_0}{a}\right)\left(1+\frac{R_I\kappa_0}{b}\right)t^2\right]\\
    &H(t)=\frac{1}{4}\left[\left(1-\frac{a^{2l}}{b^{2l}}\right)+\left(1+\frac{a}{b}\right)\left(1+\frac{a^{2l}}{b^{2l}}\right)\frac{R_I\kappa_0}{a}t\right.\nonumber\\
    &~~~~~~\left.-\left(1-\frac{a^{2l}}{b^{2l}}\right)\left(1-\frac{R_I\kappa_0}{b}\right)\left(1+\frac{R_I\kappa_0}{a}\right)t^2\right]
\end{align}
and notice that $l=\sqrt{\kappa_{\theta}/\kappa_r}=t\kappa_0/\kappa_r$ is also $t$-dependent.

To eliminate the external-field distortion, $E_1$ in Eq. (13) must be zero. This means $H(t)$ must have a root $t_0>0$. Since the first two terms in $H(t)$ are greater than zero and $H(t=1)>0$, the sufficient and necessary condition for $H(t)$ having a root $t_0>0$ is the coefficient of the second order term $t^2$ is less than zero, i.e.
\begin{equation}
    1-\frac{R_I\kappa_0}{b}>0, ~\text{or}~L_c<b
\end{equation}
and the root $t_0$ will satisfy $t_0>1$. As an example, we keep the geometry parameters $a$ and $b$, thermal conductivity $\kappa_0$ and $\kappa_r$, and ITR ($R_I$) unchanged, as those used in Fig. 3(b), so $L_c\equiv R_I\kappa_0<b$ is satisfied. The root of $H(t)$ is $t_0=1.276$, which gives $\kappa_{\theta}=t_0^2\kappa_0^2/\kappa_r=4.13$ W/(m$\cdot$K). Fig. 3(c) and 3(g) show the numerical results using this modified thermal conductivity, where the external-field distortion is eliminated. Furthermore, it is easy to find that $F(t_0)>F(t=1)$ ($t=1$ is the case of Eq. (15)), which means the heat flux in cloaking region (I) is smaller when using the modified thermal conductivity. 

If we further increase $\kappa_{\theta}$, the external field is distorted again, as shown in Fig. 3(d) and 3(h). But this time the isothermal contour lines become outward curves, opposite to the inward curves in Fig. 3(b). This means that the required $\kappa_{\theta}$ to eliminate external-field distortion is unique.

\section{Bilayer Cloak}

\begin{figure}[htbp]
    \centering
    \includegraphics[width=0.6\linewidth]{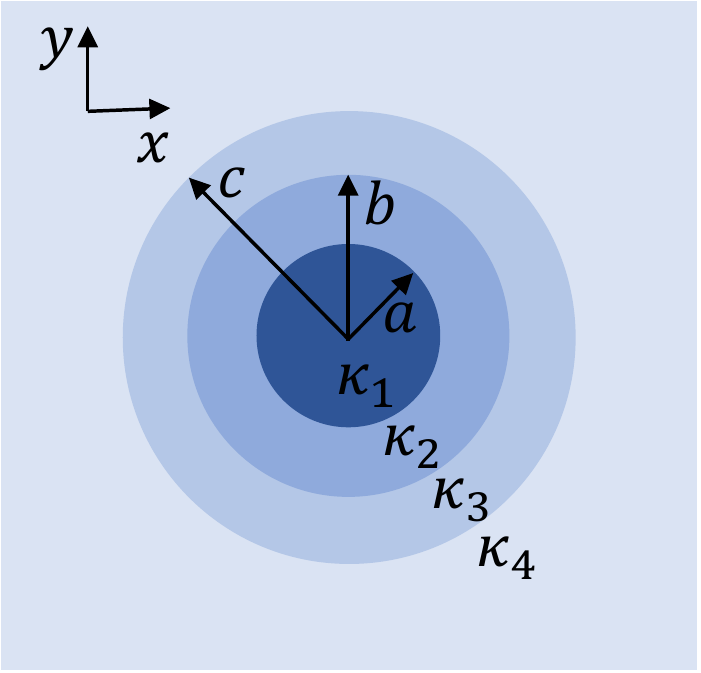}
    \caption{Schematic of bilayer thermal cloak.}
    \label{fig:structbi}
\end{figure}

Our second example is the 2D bilayer cloak structure, as shown in Fig. 5, where $\kappa_i$ $(i=1,2,3,4)$ is isotropic. The cloak consists of two annular layers with thermal conductivity $\kappa_2$ and $\kappa_3$. $\kappa_1$ is the thermal conductivity of the cloaking object and $\kappa_4$ is that of background. The whole structure is presented in a uniform thermal gradient field $\nabla T_b=G_0\hat{x}$. This thermal cloak has been experimentally realized by Han et al \cite{han2014experimental}. Similar to the anisotropic single layer case, the temperature in all the regions of space can be generally expressed as \cite{han2014experimental}
\begin{equation}
    T_i=\sum_{n=1}^{\infty}\left[A_{n}^ir^{n}+B_{n}^ir^{-n}\right]\cos{n\theta}
\end{equation}
where $T_i$ is the temperature of the $i$th layer, $A_{n}^i, B_{n}^i$ $(i=1,2,3,4)$ are constants determined by the boundary conditions Eq. (4) and 
\begin{align}
    T_1|_{r=0}~\text{is finite},\quad T_4|_{r\rightarrow\infty}=-G_0r\cos{\theta}
\end{align}

For simplicity, considering that the 
inner layer is perfect insulation material, i.e., $\kappa_2=0$
, this ensures that an external field does not 
penetrate inside the cloaking region and the only task is to eliminate the external-field distortion. Combining boundary conditions Eq. (4), (21), and the general solutions Eq. (20), we obtain 
\begin{align}
    &A_1^4=-G_0\\
    &B_1^4=-\frac{G_0 c^2 \left[(c^2-b^2)(R_{34}\kappa_4-c)\kappa_3+c(b^2+c^2)\kappa_4\right]}{(c^2-b^2)(R_{34}\kappa_4+c)\kappa_3+c(b^2+c^2)\kappa_4}
\end{align}
and all other $A_{n}^4$'s and $B_{n}^4$'s are zero. $R_{34}$ denotes the ITR between layer 3 and 4. $A_1^4$ represents the background thermal gradient field and $B_1^4$ represents the distortion in region 4 induced by the cloak. To eliminate the external-field distortion requires $B_1^4=0$, which means the numerator must be zero
\begin{equation}
    (c^2-b^2)(R_{34}\kappa_4-c)\kappa_3+c(b^2+c^2)\kappa_4=0
\end{equation}
We notice that $(c^2-b^2)>0$, $\kappa_3>0$, and $c(b^2+c^2)\kappa_4>0$, so the existence of a solution requires 
\begin{equation}
    R_{34}\kappa_4-c<0,~\text{or}~R_{34}\kappa_4<c
\end{equation}
Eq. (25) again gives a relationship between the characteristic length $L_c$ and the cloak size $c$. Under this constraint, the solution of Eq. (24) gives 
\begin{equation}
    \kappa_3=\frac{c^2+b^2}{(c^2-b^2)(1-R_{34}\kappa_4/c)}\kappa_4
\end{equation}

To verify the results above, we simulate the temperature distribution and heat flux of the bilayer structure realized in Ref. \cite{han2014experimental} using FEM, but with the cloak size $10^4$ times smaller. Fig. 6(a) shows the simulation result without ITR. If ITR is considered, the external field is significantly distorted, as shown in Fig. 6(b) and 6(f). We use Eq. (26) to eliminate the external-field distortion. Fig. 6(c) shows the temperature distribution with the modified thermal conductivity $\kappa_3$, where nearly no external-field distortion (see Fig.6(g)) is observed. If we further increase $\kappa_3$, the external field is distorted again, as shown in Fig. 6(d). But this time the isothermal contour lines become outward curves, opposite to the inward curves in Fig. 6(b). This behaviour is similar to what we find in Section II. 
\begin{figure}[htbp]
    \centering
    \includegraphics[width=0.9\linewidth]{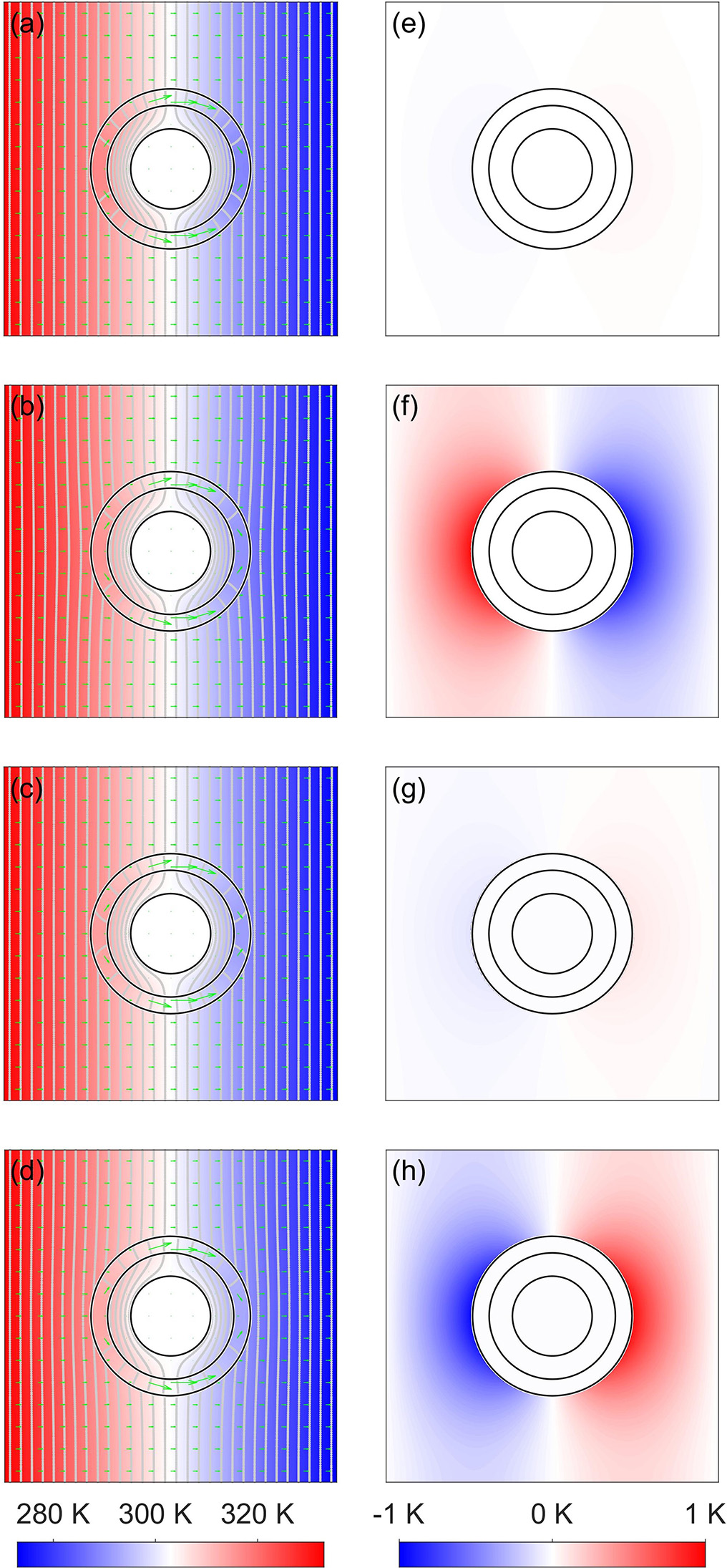}
    \caption{Temperature distribution of the bilayer structure. The green arrows show the heat flux distribution Temperature at the left (right) boundary is fixed at 333 K (273 K). $a=0.6$ $\mu$m, $b=0.95$ $\mu$m, $c=1.2$ $\mu$m, $\kappa_1=27$ W/(m$\cdot$K), $\kappa_2=0.03$ W/(m$\cdot$K), $\kappa_4=2.3$ W/(m$\cdot$K). (a) $\kappa_3=9.8$ W/(m$\cdot$K), $R_{34}=0$ K$\cdot$m$^2$/W. (b) $\kappa_3=9.8$ W/(m$\cdot$K), $R_{34}=1\times10^{-7}$ K$\cdot$m$^2$/W. (c) $\kappa_3=12.4$ W/(m$\cdot$K) determined by Eq. (29), $R_{34}=1\times10^{-7}$ K$\cdot$m$^2$/W. (d) $\kappa_3=15$ W/(m$\cdot$K), $R_{34}=1\times10^{-7}$ K$\cdot$m$^2$/W. (e)-(h) Temperature difference between the temperature distribution in (a)-(d) and uniform thermal gradient field. Only the difference outside the cloak region is calculated because we want to study the external-field distortion.}
    \label{fig:bicol}
\end{figure}

In cloaking region ($r<a$), the average heat flux is calculated via
\begin{equation}
    \overline{|\bm{q}|}=\frac{1}{\pi a^2}\iint_{r<a}|\bm{q}|dxdy
\end{equation}
to compare the shielding performance of thermal cloak. The result is $\overline{|\bm{q}|}_{(b)}/\overline{|\bm{q}|}_{(a)}=0.88$, $\overline{|\bm{q}|}_{(c)}/\overline{|\bm{q}|}_{(a)}=0.79$, where the subscript (a), (b), (c) denotes the case of Fig. 6(a), 6(b), 6(c) respectively. This indicates the reduction of heat flux when considering ITR.

\begin{figure}[htbp]
    \centering
    \includegraphics[width=0.5\linewidth]{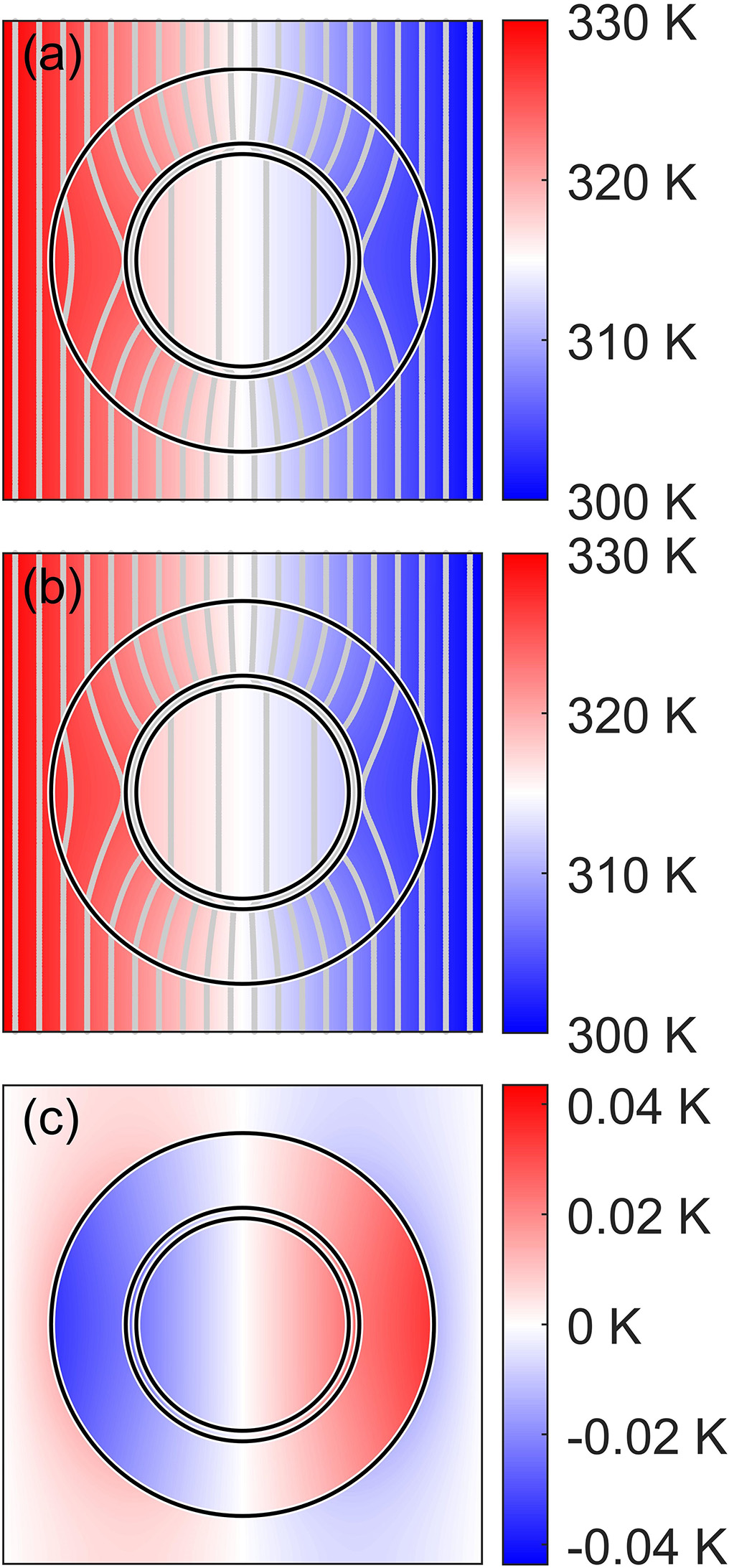}
    \caption{Temperature distribution of the thermal cloak realized in Ref. \cite{choe2019ion}. $a=5.55$ $\mu$m, $b=6.1$ $\mu$m, $c=10$ $\mu$m, $\kappa_1=\kappa_4=40$ W/(m$\cdot$K), $\kappa_2=2$ W/(m$\cdot$K), $\kappa_3=65$ W/(m$\cdot$K). (a) without ITR. (b) with ITR. (c) Temperature difference ($T_b-T_a$) between (a) and (b).}
    \label{fig:nano}
\end{figure}
Next, we further consider the experiment in Ref. \cite{choe2019ion}, where they use crystalline and amorphous silicon to realize thermal cloak at microscale. ITR between crystalline and amorphous silicon at 300 K is about $1\times 10^{-9}$ K$\cdot$m$^2$/W \cite{france2014thermal}. As shown in Fig. 7, ITR only introduces very small difference to the temperature distribution, which means it is a good approximation to ignore ITR in their experiment. This is easy to understand since the characteristic length $L_c=R_{I}\kappa_{4}\approx 4\times10^{-8}$ m is much less than the cloak size $c=10$ $\mu$m ($10^{-5}$ m). But if the cloak size decreases to the order of 0.1 $\mu$m ($10^{-7}$ m), i.e. 100 times smaller, the effect of ITR will become significant.

\section{Conclusion and discussions}
 In this paper, we have studied the effect of ITR in steady thermal cloak. In the two cloaks we investigate, ITR improves the performance of thermal shielding while destroys the invisibility of thermal cloak. We have introduced a characteristic length, which is the product of the ITR and background thermal conductivity, to quantify the effect of ITR. The effect becomes significant when the characteristic length is comparable to the cloak size. If the characteristic length is smaller than the cloak size, we can modify the thermal conductivity to restore the invisibility feature. Although the structures we have studied are simple, it is straightforward to generalize our analysis to more complex multilayer structures. We believe the improvement in thermal shielding is generic, and to restore the invisibility feature is still possible if the characteristic length is smaller than the cloak size.
 
In non-steady thermal cloak, Sklan et al \cite{sklan2016detecting} have demonstrated that the bilayer cloak can be detected by transient temperature distribution. We also investigate the thermal cloak with ITR and modified thermal conductivity, and find almost the same transient effect, which means ITR cannot be used to remove the transient effect.


%

\end{document}